 \documentclass[smallabstract,smallcaptions]{dccpaper}

\usepackage{epsfig}
\usepackage{amsmath}
\usepackage{amssymb}
\usepackage{color}
\usepackage{url}
\usepackage{bm}
\usepackage{subfigure}
\usepackage{stfloats}
\usepackage{latexsym}
\usepackage{float}
\usepackage{psfrag}
\usepackage{cite}
\usepackage{textcomp}
\usepackage{graphicx,times}
\usepackage{epsfig,latexsym}
\usepackage{float}
\usepackage{indentfirst}
\usepackage{amsmath}
\usepackage{amssymb}
\usepackage{times}
\usepackage{psfrag}
\usepackage{cite}
\usepackage{multirow}
\usepackage{subfigure}
\usepackage{rotating}
\usepackage{graphicx}
\usepackage{epstopdf}

\newlength{\figurewidth}
\newlength{\smallfigurewidth}

\begin{document}

\newcommand{\argmin}{\operatornamewithlimits{argmin}}

\title
{\large
\textbf{Watching Videos with Certain and Constant Quality: PID-based Quality Control Method}
}

\author{%
Yuhang Song$^{1}$, Mai Xu$^{1~*}$ and Shengxi Li$^{1}$\\[0.5em]
{\small\begin{minipage}{\linewidth}\begin{center}
\begin{tabular}{ccc}
$^{1}$Beihang University & \hspace*{0.5in} \\
37 Xueyuan Road \\
Beijing, 100191, China \\
\url{MaiXu@buaa.edu.cn}
\end{tabular}
\end{center}\end{minipage}}
}

\maketitle
\thispagestyle{empty}

\begin{abstract}

In video coding, compressed videos with certain and constant quality can ensure quality of experience (QoE). To this end, we propose in this paper a novel PID-based quality control (PQC) method for video coding. Specifically, a formulation is modelled to control quality of video coding with two objectives: minimizing control error and quality fluctuation. Then, we apply the Laplace domain analysis to model the relationship between quantization parameter (QP) and control error in this formulation. Given the relationship between QP and control error, we propose a solution to the PQC formulation, such that videos can be compressed at certain and constant quality. Finally, experimental results show that our PQC method is effective in both control accuracy and quality fluctuation.
\end{abstract}
\vspace{-1.0em}
\section{Introduction}
\vspace{-1.0em}
Along with the explosive increase of multimedia content, effective and efficient compression algorithms have been always in demand for decades. Catering for such demand, various video coding standards have been proposed and developed to compress raw video data \cite{wiegand2003overview, sullivan2012overview, mukherjee2013latest}, in the way of minimizing bit-rates or optimizing quality with certain constraints according to different applications. Generally speaking, rate control \cite{li2014domain, li2016opti} aims at controlling bit-rates to meet different requirements, e.g., minimizing distortion for storage application or reducing bit-rates fluctuation for communication. Moreover, in some cases when perceived quality is highly crucial, quality control can be adopted to compress a video at a certain and constant quality, thus obtaining more desirable QoE.

\begin{figure}[h]
  \begin{center}
   \includegraphics[width=1.0\linewidth]{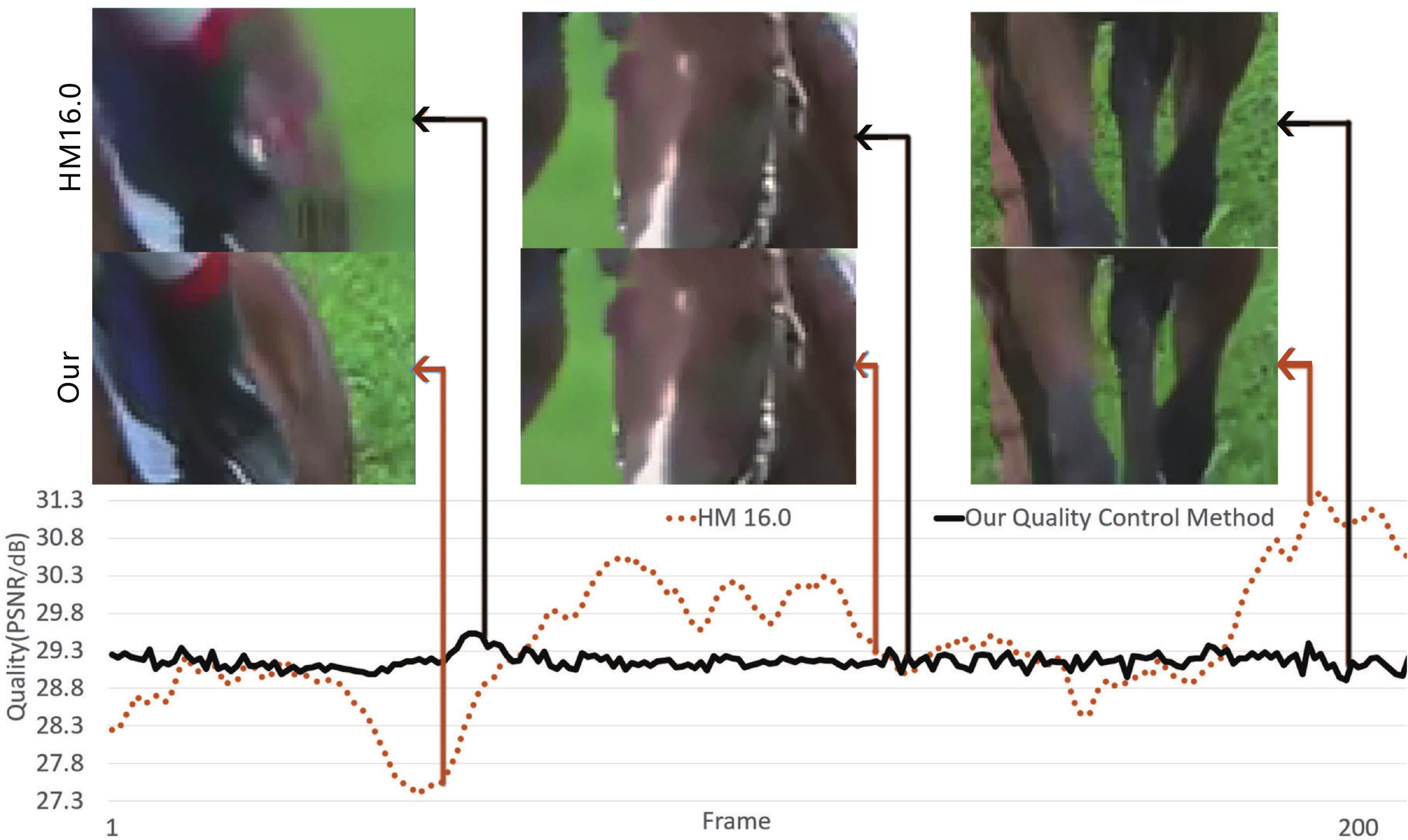}
  \end{center}
  \vspace{-1.5em}
\caption{\label{fig1}\footnotesize{An example of quality control in video coding. In this figure, the state-of-the-art video coding standard high efficiency video coding (HEVC), which is implemented by the latest reference software HM 16.0, is used as an anchor. This figure shows that the fluctuation of the HEVC encoded video may result in poor perceived quality for some frames (e.g., frames 55, 120 and 190). By contrast, our quality control method is able to yield smooth quality, better than the poor quality of frames 55, 120 and 190 in HEVC. Besides, our method ensures distortion of encoded video approaching to the target, i.e., 29.2 dB.}}
\end{figure}

Similar to constant rate control which aims at smoothing bit-rates for each frame to avoid buffer overflow or underflow, constant quality control (see Figure \ref{fig1}) in video coding provides smooth quality in compressed frames to avoid overall perceived quality degradation caused by some intense quality fluctuation. Toward this end, a direct way is to adopt a two-pass procedure, which pre-analyzes all the frames in advance and calculates the global quantization parameters (QPs) for each frame \cite{yu2001novel} \cite{lin1998bit} \cite{lee1994temporally}. In addition, in some cases the calculated QPs are not integers, and \cite{he2001low} thus proposed a solution by adjusting QPs at block level to reach the desired QP value at frame level. Obviously, the pre-analysis strategy is not applicable to real-time streaming applications, e.g., video conferencing and live video streaming.

Later, He \emph{et.al.} \cite{he2005low} proposed a low-pass filter with geometric factor, and then applied it in quality control to achieve smoothed frame quality, meanwhile satisfying buffer status with low delay. Besides, a PSNR adjustment method was proposed in \cite{de2005psnr} to maintain the group of picture (GOP) level quality constant, by empirically moving up/down QPs according to previous frames. Instead of the GOP-based control, \cite{xie2006sequence} proposed a sequence-based method, which tracks scene changes and then achieves smooth video quality adaptive to scene changes. Recently, a trellis-based method has been proposed in \cite{huang2009consistent} to consider both bit budget and PSNR variance. Although above works can reduce quality fluctuation with constant quality, they are unable to control quality to a certain level that users demand.

Most recently, Hu \emph{et.al.} \cite{hu2012adaptive} have proposed to control the quality of encoded videos to a certain target for H.264. In their method, QP of each frame is adjusted by modelling consecutive frames, in the assumption of Laplacian distribution on transform coefficients. Based on the same assumption, \cite{seo2013rate} proposed a mixture model on Laplacian function to tailor distortion-quantization (D-Q) and rate-quantization (R-Q) relationships in HEVC, thus controlling quality in a constant manner. However, the certain assumptions, like Laplacian distribution of transform coefficients, are not precise for modelling various-content videos, thus leading to high inaccuracy in quality control. More importantly, those works are restricted to some specific standards or certain assumptions, and they are not effective when being applied to the latest R-$\lambda$ model \cite{li2016optimal} of HEVC.

In this paper, we propose a novel quality control method for video coding, which does not rely on some specific encoders or assumptions. As such, our method is suitable for different encoders in high control accuracy. To our best knowledge, our method is the first work on encoder-free quality control, which enables to reach certain and constant distortion across frames of encoded videos. Specifically, our method is inspired by the basic idea of the proportion-integral-derivative (PID) controller, thus called the PID-based quality control (PQC) method. We first propose a formulation of quality control in video coding for our PQC method, with objectives of minimization on both control error and fluctuation of distortion. Then, the relationship between QP and distortion is modelled for the formulation of our PQC method, such that the PID controller can be used to solve our formulation. Next, a PID-based solution is provided for our quality control formulation with modelled relationship between $\mathrm{QP}$ and distortion. As such, the distortion of video coding can be controlled to a target distortion with small quality fluctuation. Finally, we implement our PQC method in the latest high efficiency video coding (HEVC) encoder (HM 16.0), and our experimental results show that our method achieves the state-of-the-art quality control performance in terms of both control error and quality fluctuation.

\vspace{-1.0em}
\section{Overview of PID Controller}\label{pid-overview}
\vspace{-1.0em}
The PID controller\cite{ang2005pid} is widely used to minimize the error between the measured step point and the target step point. Using the terms of proportional operator ($P$), integral operator ($I$) and derivative operator ($D$), a PID controller maintains a well trade-off among response speed, static error correction and overreacting repression. Generally speaking, the PID controller performs robustly with little overhead of computational complexity. Thus, our quality control method for video coding is based on the PID controller.

To be more specific, assume that there exists error between the current and target positions until time $(t-1)$, which are denoted by $\lbrace e_{t-1},e_{t-2},\cdots,e_{0} \rbrace$. The PID controller focuses on minimizing error $e_t$ by adjusting a control variable $o_t$. In the PID controller, $o_t$ can be calculated by
\vspace{-0.7em}
\begin{equation}
\label{PID-equation}
o_t=K_pe_{t-1}+K_i\int^{t-1}_0e_{\tau}d{\tau}-K_d\dfrac{de_{t-1}}{dt},
\vspace{-0.7em}
\end{equation}
where $e_{t-1}$, $\int^{t-1}_0e_{\tau}d{\tau}$ and $\dfrac{de_{t-1}}{dt}$ are the proportion ($P$), integral ($I$) and derivative ($D$) values; $K_p$, $K_i$ and $K_d$ (all $\geq$ 0) denote their corresponding weights. As can be seen from \eqref{PID-equation}, $P$ is decided by the most recent error, whilst $I$ accounts for long lasting previous error and $D$ predicts error in the future. For more details about the setting of $K_p$, $K_i$ and $K_d$, refer to \cite{cominos2002pid}. To sum up, the PID controller outputs an optimal predicted value of the control variable to minimize error $e_t$, using all the error incurred until time $(t-1)$. Next, we propose to control quality of video coding by incorporating the PID controller.
\vspace{-1.0em}
\section{PID-based quality control method}\label{PID-based-quality-control-method}
\vspace{-1.0em}
In this section, we present our encoder-free PQC method for video coding, which is achieved by predicting the optimal $\mathrm{QP}$ before encoding each frame. Specifically, we first establish in Section \ref{Framework of PID-based-quality-control-method} the formulation  of quality control in video coding. In Section \ref{relationship-qp-e}, we further model the relationship between $\mathrm{QP}_{t}$ and $e_t$ for the proposed quality control formulation. In Section \ref{Policy Function with PID Controller}, we solve the proposed quality control formualtion using the PID controller.
\vspace{-1.0em}
\subsection{Formulation of quality control}\label{Framework of PID-based-quality-control-method}
\vspace{-0.5em}
In video coding, there are two main objectives for quality control:

$\textbf{Objective I}$: Minimizing the error between the actual and target quality, averaged over all frames.

$\textbf{Objective II}$: Minimizing the fluctuation of quality along with frames.

The above two objectives can be achieved by predicting the optimal $\mathrm{QP}$ before encoding each frame. In other words, before encoding the $t$-th frame, we need to estimate the best $\mathrm{QP}$ value for this frame, which is denoted by $\mathrm{QP}_{t}$. Assuming that $T$ is the target distortion and $D_t$ is the distortion of the $t$-th frame, the quality control can be formulated by
\vspace{-0.7em}
\begin{equation}
\label{quality-control-formulation}
\mathrm{QP}_{t}=\argmin_{\mathrm{QP}} \lbrace \underbrace{\lambda \cdot \underbrace{(D_{t}(\mathrm{QP})-T)}_{\textbf{Objective I}}+(1-\lambda) \cdot \underbrace{\dfrac{dD_{t}(\mathrm{QP})}{dt}}_{\textbf{Objective II}}}_{e_t} \rbrace,
\vspace{-0.7em}
\end{equation}
where $(D_{t}(\mathrm{QP})-T)$ modells the error between the actual and target quality ($\textbf{Objective I}$), while $\frac{dD_{t}(\mathrm{QP})}{d_t}$ modells the fluctuation of quality ($\textbf{Objective II}$). In addition, $\lambda$ represents the trade-off between the two objectives, and $e_t$ denotes the overall error to be minimized.

Since PSNR is a widely used distortion evaluation metric for video coding, it is applied to model distortion $D_t$ in \eqref{quality-control-formulation} for this paper. Moreover, our PQC method can be simply extended to adopting other distortion evaluation metrics, like Structural Similarity Index (SSIM) and Video Signal-to-Noise Ratio (VSNR). To solve formulation \eqref{quality-control-formulation}, we first need to model the relationship between $\mathrm{QP}_{t}$ and $e_t$, which is to be discussed in the following.
\vspace{-1.0em}
\subsection{Relationship between $\mathrm{QP}_{t}$ and $e_t$}\label{relationship-qp-e}
\vspace{-0.5em}
In a coding system, the relationship between $\mathrm{QP_t}$ and $e_t$ can be modelled by the following function $\Psi$,
\begin{equation}
\label{coding-system}
\Psi(\mathbf{I}_{t},\mathbf{I}_{t-1},\cdots,\mathbf{I}_0,\mathrm{QP}_{t},\mathrm{QP}_{t-1},\cdots,\mathrm{QP}_{0})=e_t,
\vspace{-0.5em}
\end{equation}
where $\mathbf{I}_t$ stands for the frame content at frame $t$. This formulation shows that content and $\mathrm{QP}$s of all frames until currently encoded frame contribute to quality control error $e_t$, for a given encoder. In fact, $\mathbf{I}_{t},\mathbf{I}_{t-1},\cdots,\mathbf{I}_0$ is a set of images from the video sequence. We denote them by a single tensor $\mathbb{I}$. Our intention here is to analyze the relationship between $\mathrm{QP}_{t}$ and $e_t$. Thus, given a sequence, we have a fixed $\mathbb{I}$, and then \eqref{coding-system} can be rewritten by
\begin{equation}
\label{coding-system-i-down}
\Psi_\mathbb{I}(\mathrm{QP}_{t},\mathrm{QP}_{t-1},\cdots,\mathrm{QP}_{0})=e_t.
\vspace{-0.5em}
\end{equation}
Obviously, $\Psi_\mathbb{I}$ describes the relationship between $\mathrm{QP}_{t}$ and $e_t$ for given $\mathbb{I}$. To obtain $\Psi_\mathbb{I}$, we propose a simple and practical way from the viewpoint of signal processing by treating $\Psi_\mathbb{I}$ as an unknown linear time invariant (LTI) system (Input: $\mathrm{QP}_{t}$ and Output: $e_t$).

\begin{figure}[t]
\begin{center}
\begin{tabular}{cc}
\multicolumn{2}{c}{} \\[-1em]
\subfigure{\includegraphics[width=3.0in]{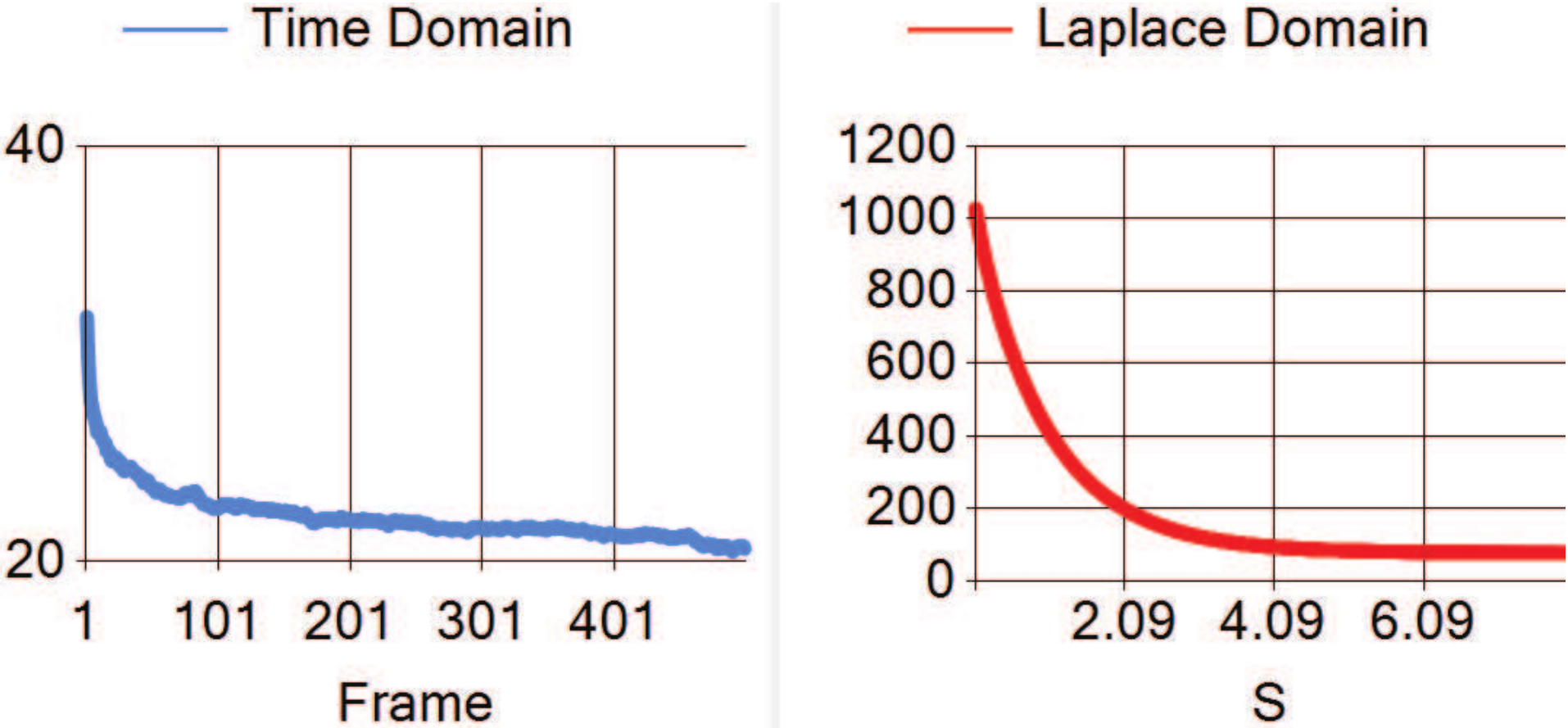}} &
\subfigure{\includegraphics[width=3.0in]{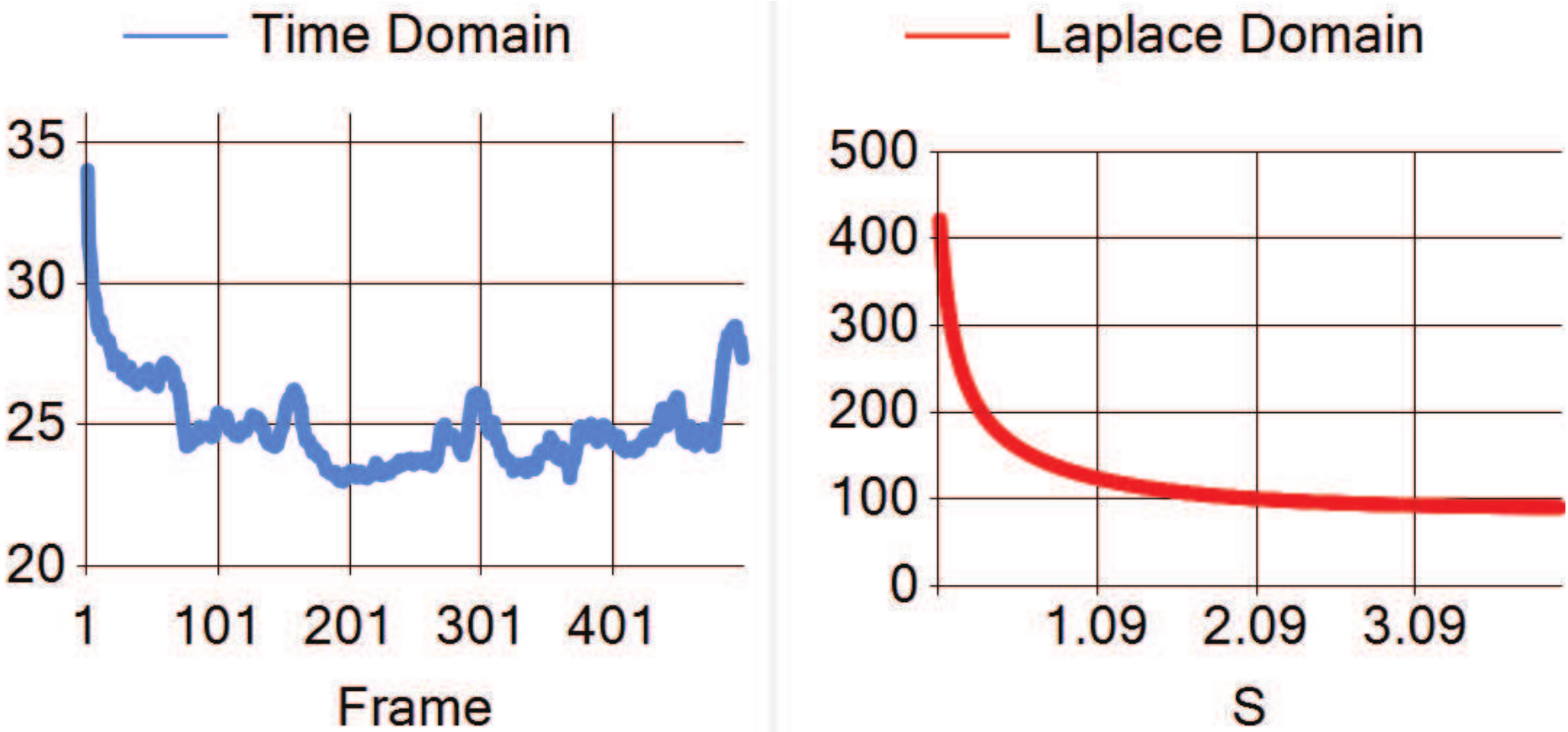}} \\
{\small \footnotesize{BasketballDrill (inter coding)}} & {\small \footnotesize{BasketballPass (inter coding)}}\\
\multicolumn{2}{c}{} \\[-0.5em]
\subfigure{\includegraphics[width=3.0in]{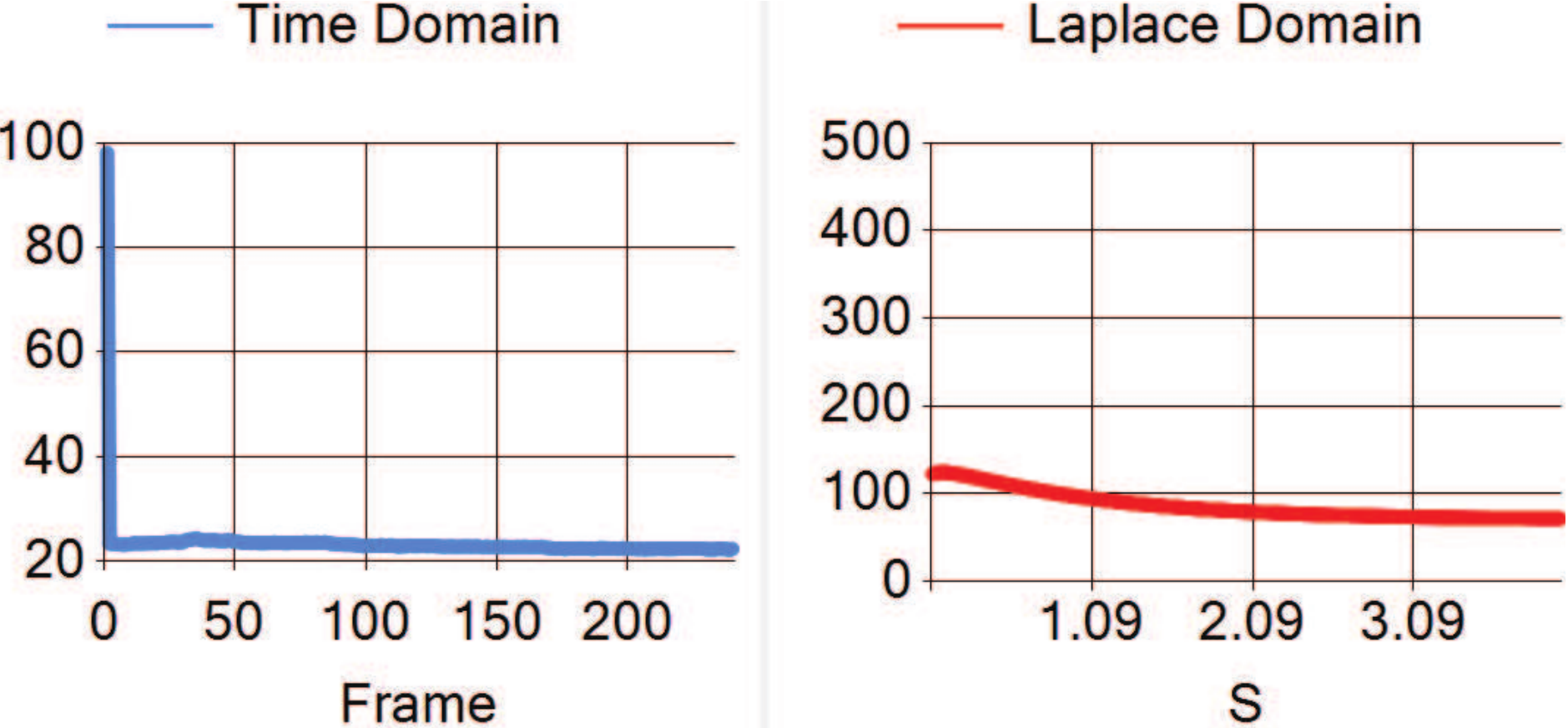}} &
\subfigure{\includegraphics[width=3.0in]{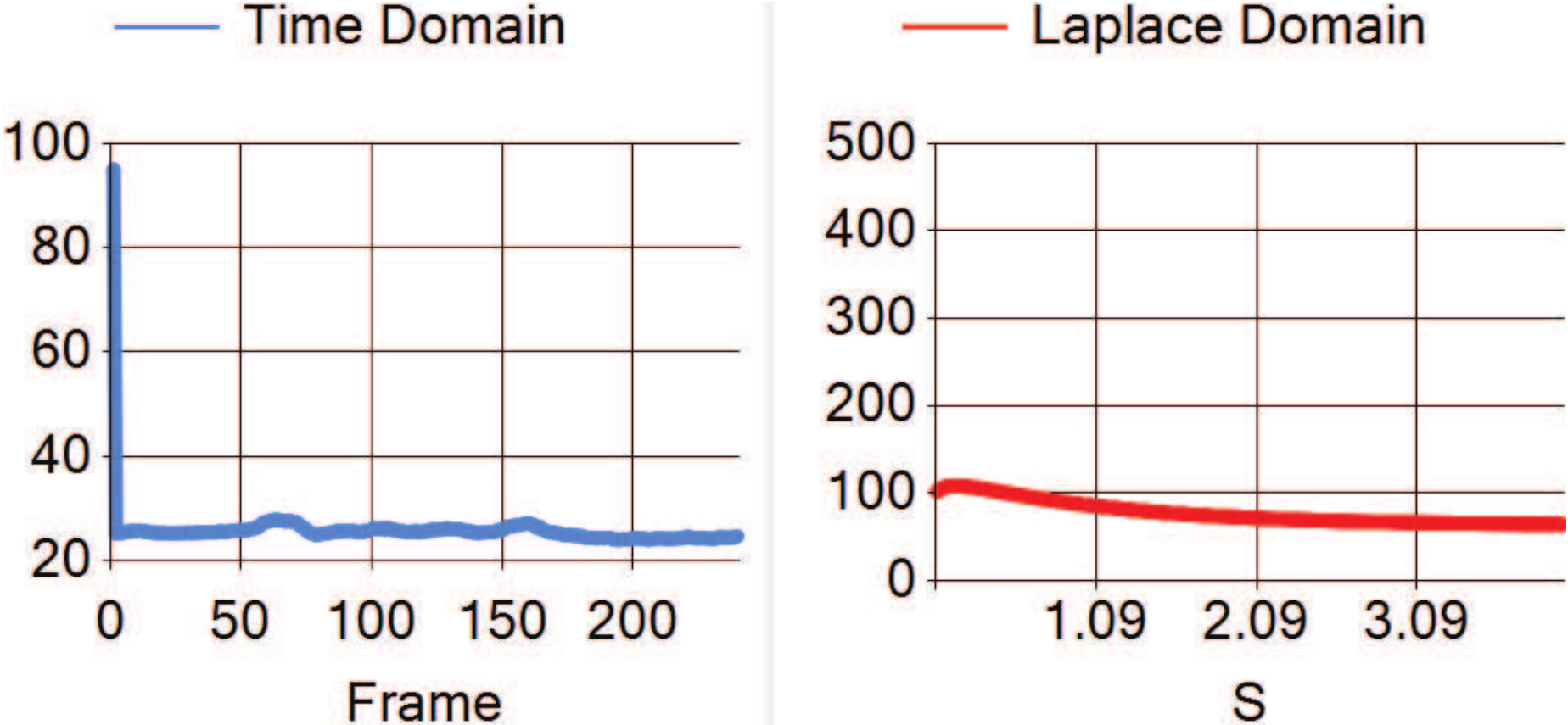}} \\
{\small \footnotesize{BlowingBubbles (intra coding)}} & {\small \footnotesize{BasketballPass (intra coding)}}\\
\end{tabular}
\end{center}
\vspace{-1.5em}
\caption{\label{anay}%
\footnotesize{Analysis of practical $\Psi_\mathbb{I}(\mathbf{QP})$ upon the state-of-the-art HEVC encoder with inter and intra coding. For analyzing $\Psi_\mathbb{I}(\mathbf{QP})$, we encoded 2 sequences by HM 16.0, using the configuration file \emph{encoder$\_$lowdelay$\_$P$\_$main.cfg} and \emph{encoder$\_$intra$\_$main.cfg}. Here, the frames of each sequence were encoded with $\mathbf{QP}=\lbrace \mathrm{QP}_{0}=\mathrm{QP}_{\mathrm{min}}, \mathrm{QP}_{1}=\mathrm{QP}_{\mathrm{max}}, \mathrm{QP}_{2}=\mathrm{QP}_{\mathrm{max}},\cdots,\mathrm{QP}_{t}=\mathrm{QP}_{\mathrm{max}} \rbrace$, in which $\mathrm{QP}_{\mathrm{min}}=0$ and $\mathrm{QP}_{\mathrm{min}}=51$ for HEVC. Then, we plot $\lbrace e_1,e_2, \cdots ,e_t \rbrace$ of each encoded frame as the ``Time Domain curve''. We further process Laplace transform on ``Time Domain curve'' of $\Psi_\mathbb{I}(\mathbf{QP})$ to obtain ``Laplace Domain curve''. It is obvious that ``Laplace Domain curve'' has 1 and 0 pole for inter and intra coding, respectively.}}
\end{figure}


First, the encoding system is input with an impulse signal $\mathbf{QP}=\lbrace \mathrm{QP}_{0}=\mathrm{QP}_{\mathrm{min}}, \mathrm{QP}_{1}=\mathrm{QP}_{\mathrm{max}}, \mathrm{QP}_{2}=\mathrm{QP}_{\mathrm{max}},\cdots,\mathrm{QP}_{t}=\mathrm{QP}_{\mathrm{max}} \rbrace$, where $\mathrm{QP}_{\mathrm{min}}$ and $\mathrm{QP}_{\mathrm{max}}$ are possibly minimal and maximal $\mathrm{QP}$ values of the encoder, respectively. Then, the system response $\Psi_\mathbb{I}(\mathbf{QP})$ of each video sequence is obtained for a specific video encoder (i.e., HEVC). Next, we apply Laplace transform on the obtained $\Psi_\mathbb{I}(\mathbf{QP})$ to estimate the number of poles on its Laplace domain. The results are shown in Figure \ref{anay}. From this figure, we can see that $\Psi_\mathbb{I}(\mathbf{QP})$ has 1 and 0 pole on Laplace domain for inter frame and intra frame encoding, respectively. As such, $\Psi_\mathbb{I}(\mathbf{QP})$ can be seen as first- and zero- order systems for inter and intra frame video coding \cite{widder2015laplace}\cite{dorf1998modern}. Therefore, $\Psi_\mathbb{I}(\mathbf{QP})$ can be represented by the differential equations, as follows,

\begin{equation}
\label{inter-relation}
\mathbf{Inter~Frame:~}A_1^\mathbb{I} \cdot e_t + A_2^\mathbb{I} \cdot \frac{de_t}{dt} = \mathrm{QP}_t,
\end{equation}
\begin{equation}
\label{intra-relation}
\mathbf{Intra~Frame:~}A_0^\mathbb{I} \cdot e_t = \mathrm{QP}_t,
\end{equation}
where $A_0^\mathbb{I}$, $A_1^\mathbb{I}$ and $A_2^\mathbb{I}$ are the coefficients derived from the data analysis in Figure~\ref{anay}. Note that the exact values for $A_0^\mathbb{I}$, $A_1^\mathbb{I}$ and $A_2^\mathbb{I}$ are unnecessary, since our PQC method only requires the number of orders for the differential equations, when applying the PID controller to solve our quality control formulation of \eqref{quality-control-formulation}.

In following section, we focus on our solution to the proposed quality control formulation of \eqref{quality-control-formulation} on the basis of $e_t$ and $\mathrm{QP}_{t}$ relationship of \eqref{inter-relation} \eqref{intra-relation} and the PID controller of \eqref{PID-equation}.
\vspace{-1.0em}
\subsection{PID-based solution to the quality control formulation}\label{Policy Function with PID Controller}
\begin{figure}
  \begin{center}
   \includegraphics[width=1.0\linewidth]{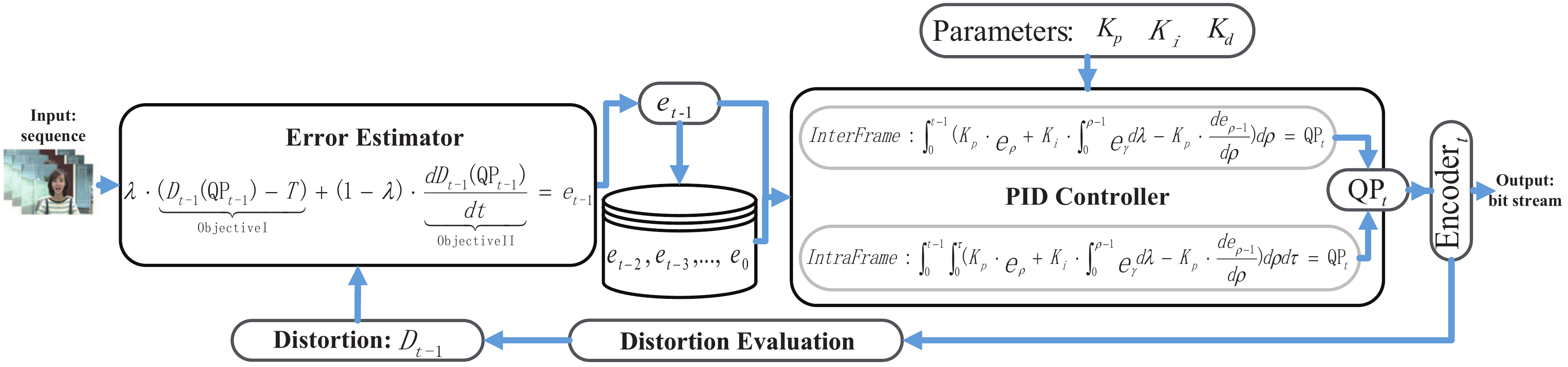}
  \end{center}
  \vspace{-1.5em}
\caption{\label{pqc-overview}%
Framework of our PQC method.}
\vspace{-1em}
\end{figure}

\vspace{-0.5em}
In this section, we apply the PID controller to solve \eqref{quality-control-formulation}. As mentioned in Section \ref{pid-overview}, the PID controller minimizes error $e_t$ alongside time $t$, via adjusting control variable $o_t$. However, the PID controller can perform well only when applied to a second-order system \cite{zhao2005fractional}. In other words, $e_t$ and $o_t$ in \eqref{PID-equation} need to satisfy the following differential equation:

\begin{equation}
\label{pid-controller-satisfy}
M_2 \cdot \frac{d^2e_t}{dt^2} + M_1 \cdot \frac{de_t}{dt} + M_0 \cdot e_t = o_t,
\end{equation}
where $M_2$, $M_1$ and $M_0$ are coefficients. Next, we use the following way to make the modelled relationship between $e_t$ and $\mathrm{QP}_{t}$ (i.e., \eqref{inter-relation} and \eqref{intra-relation}) satisfy the above requirement of the PID controller (i.e., \eqref{pid-controller-satisfy}), such that, the PID controller can be applied to solve our quality control formulation in Section \ref{Framework of PID-based-quality-control-method}. Specifically, by applying the differential operation on \eqref{inter-relation}, the following equation holds:
\begin{equation}
\label{weifen-tui}
A_1^\mathbb{I} \cdot \frac{d^2e_t}{dt^2} + A_0^\mathbb{I} \cdot \frac{de_t}{dt} +0 \cdot e_t= \frac{d\mathrm{QP}_t}{dt}.
\end{equation}
Thus, we can see that \eqref{weifen-tui} meets the requirement \eqref{pid-controller-satisfy} of the PID controller with $M_2=A_1^\mathbb{I}$, $M_1=A_0^\mathbb{I}$ and $M_0=0$. As a result, we have
\begin{equation}
\label{ddd}
\frac{d\mathrm{QP}_t}{dt}=o_t.
\end{equation}
Then, \eqref{ddd} can be rewritten as follows,
\begin{equation}
\mathbf{Inter~Frame:~}\mathrm{QP}_{t}=\int^t_0 o_\tau d\tau.
\end{equation}
Similarly, on the basis of \eqref{intra-relation} and \eqref{pid-controller-satisfy}, we have the following equation for the intra frames of video coding:
\begin{equation}
\mathbf{Intra~Frame:~}\mathrm{QP}_{t}=\int_0^{t}\int_0^{\tau} o_\rho d{\rho}d{\tau}.
\end{equation}
Finally, by replacing $o_t$ with \eqref{PID-equation}, the $\mathrm{QP}$ values of each frame can be estimated as follows for controlling quality of video coding.
\begin{equation}
\label{final-policy-function-inter}
\mathbf{Inter~Frame:~}\mathrm{QP}_{t}=\int_0^{t} o_\tau d{\tau}=\int_0^{t-1}(K_pe_{\rho-1}+K_i\int^{\rho-1}_0e_{\gamma}d\gamma-K_d\dfrac{de_{\rho-1}}{d{\rho}})d{\rho},
\end{equation}
\begin{equation}
\label{final-policy-function-intra}
\mathbf{Intra~Frame:~}\mathrm{QP}_{t}=\int_0^{t}\int_0^{\tau} o_\rho d{\rho}d{\tau} =\int_0^{t-1}\int_0^{\tau}(K_pe_{\rho-1}+K_i\int^{\rho-1}_0e_{\gamma}d\gamma-K_d\dfrac{de_{\rho-1}}{d{\rho}})d{\rho}d{\tau},
\end{equation}
Obviously, we can see that $\mathrm{QP}_t$ is only related to the control error $\lbrace e_0,e_1,\cdots,e_{t-1} \rbrace$ and parameters $K_p$, $K_i$ and $K_d$. Thus, that our method can be seen as an encoder-free quality control method. Figure \ref{pqc-overview} summarizes the overall framework of our PQC method.
\vspace{-1em}

\section{Experimental results}
\vspace{-1.0em}
In this section, experimental results are presented to verify the effectiveness of our PQC method. Details about the parameter setting are presented in Section \ref{param-setting}. In Section \ref{perf-eva}, we evaluate the performance of our method, by comparing with \cite{seo2013rate}.
\vspace{-1.0em}
\subsection{Parameter Setting}\label{param-setting}
\vspace{-0.5em}
In our experiment, all 16 video sequences from the standard test sequence database \cite{jctvcg1200} were utilized for comparison. Moreover, we implemented our PQC method on the latest HM 16.0 platform to compress all test sequences. At the same time, the state-of-the-art quality control method \cite{seo2013rate} was utilized for comparison, which was also implemented on HM 16.0 for fair comparison. It is worth pointing out that our PQC method can be applied to other encoders, as our PQC method only depends on $\mathrm{QP}$ and visual quality. For both methods, the low delay IPPP structure was chosen, using the configuration file \emph{encoder$\_$lowdelay$\_$P$\_$main.cfg}. Two $\mathrm{QP}$ values, i.e., $\mathrm{QP}=32$ and $\mathrm{QP}=37$, were chosen in our experiments, and other parameters were set by default. Specifically, the default HM 16.0 was used to compress all 16 sequences at $\mathrm{QP}=32$ and $\mathrm{QP}=37$. Then, the distortion for each compressed sequences was obtained in terms of Y-PSNR, and then it was set as the target quality for both our and \cite{seo2013rate} methods.

For the parameters related to our PQC method, we followed the PID parameter-tuning method in \cite{cominos2002pid} to find proper values of $K_p$, $K_i$ and $K_d$. As a result, we set $K_p=2.12$, $K_i=0.10$, $K_d=0.60$. Next, we empirically set $\lambda=0.8$ to balance error and fluctuation. However, it is modifiable according to different applications.


\vspace{-1.0em}
\subsection{Performance Evaluation}\label{perf-eva}
\vspace{-0.5em}
\textbf{Quality~Control~Performance.} Control error and quality fluctuation are two crucial objectives emphasized in this paper. Thus, we compare the quality control performance between our PQC and the conventional \cite{seo2013rate} methods, in terms of control error and quality fluctuation alongside frames. As can be seen from Table \ref{result}, the 4-5th and 10-11th columns report the control error of our and \cite{seo2013rate} methods, which is the difference between actual and target Y-PSNR. It is obvious that that our method significantly reduces the control error when compared with \cite{seo2013rate}, in a magnitude of ${10}^{2}$.

\begin{figure}[!h]
\begin{center}
\begin{tabular}{ccc}
\multicolumn{2}{c}{} \\[1em]
\subfigure{\includegraphics[width=2.8in]{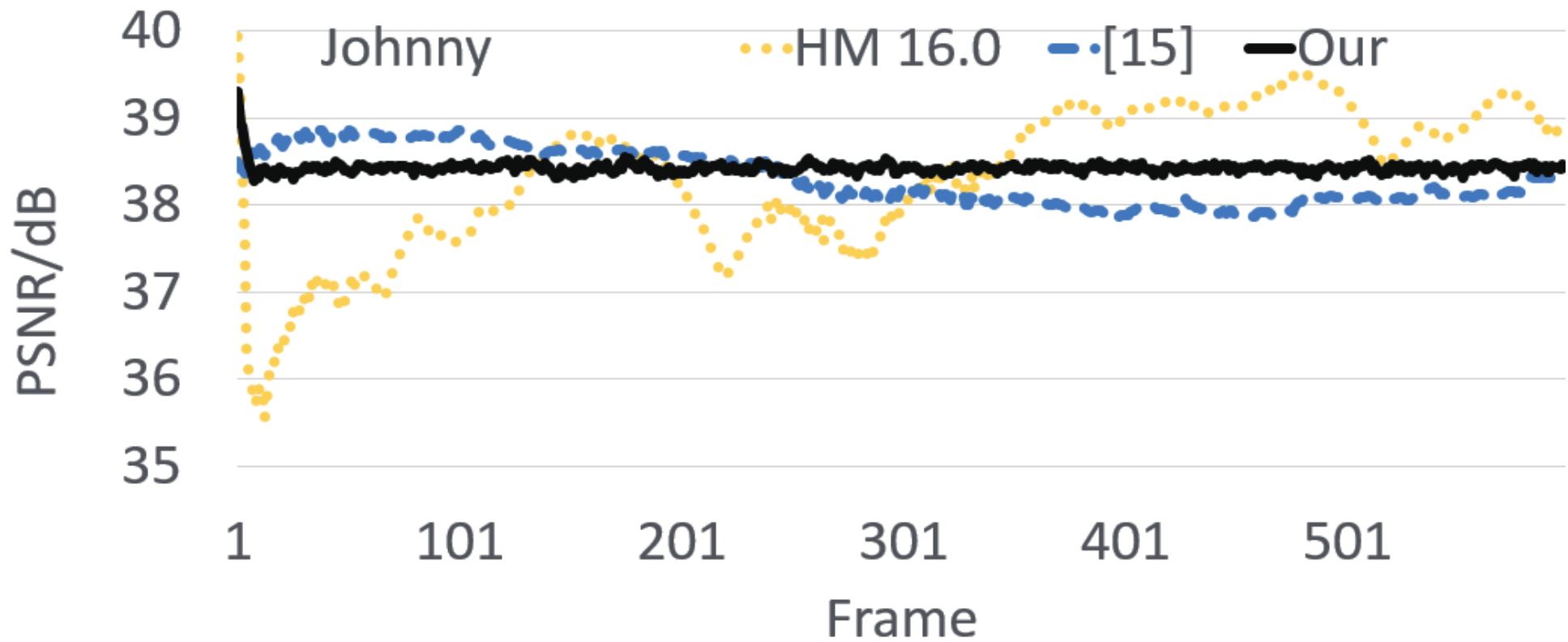}} &
\subfigure{\includegraphics[width=2.8in]{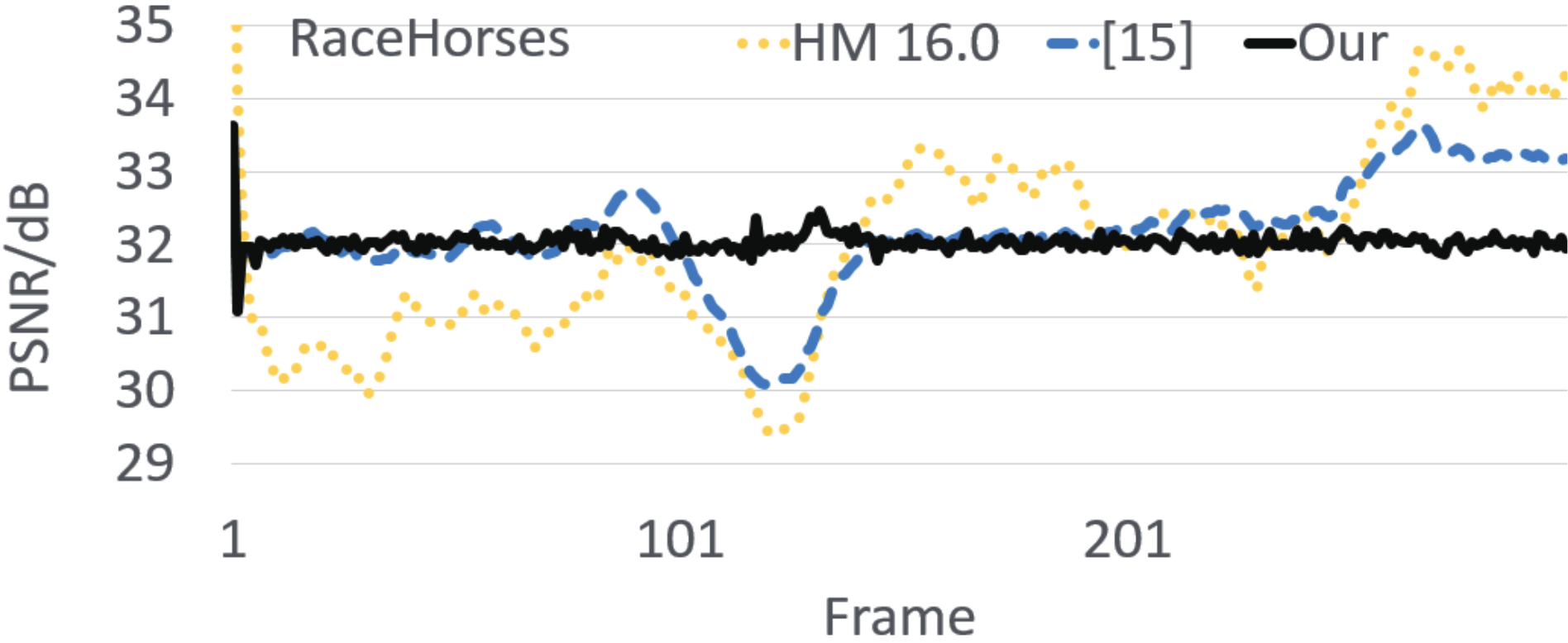}} \\
\subfigure{\includegraphics[width=2.8in]{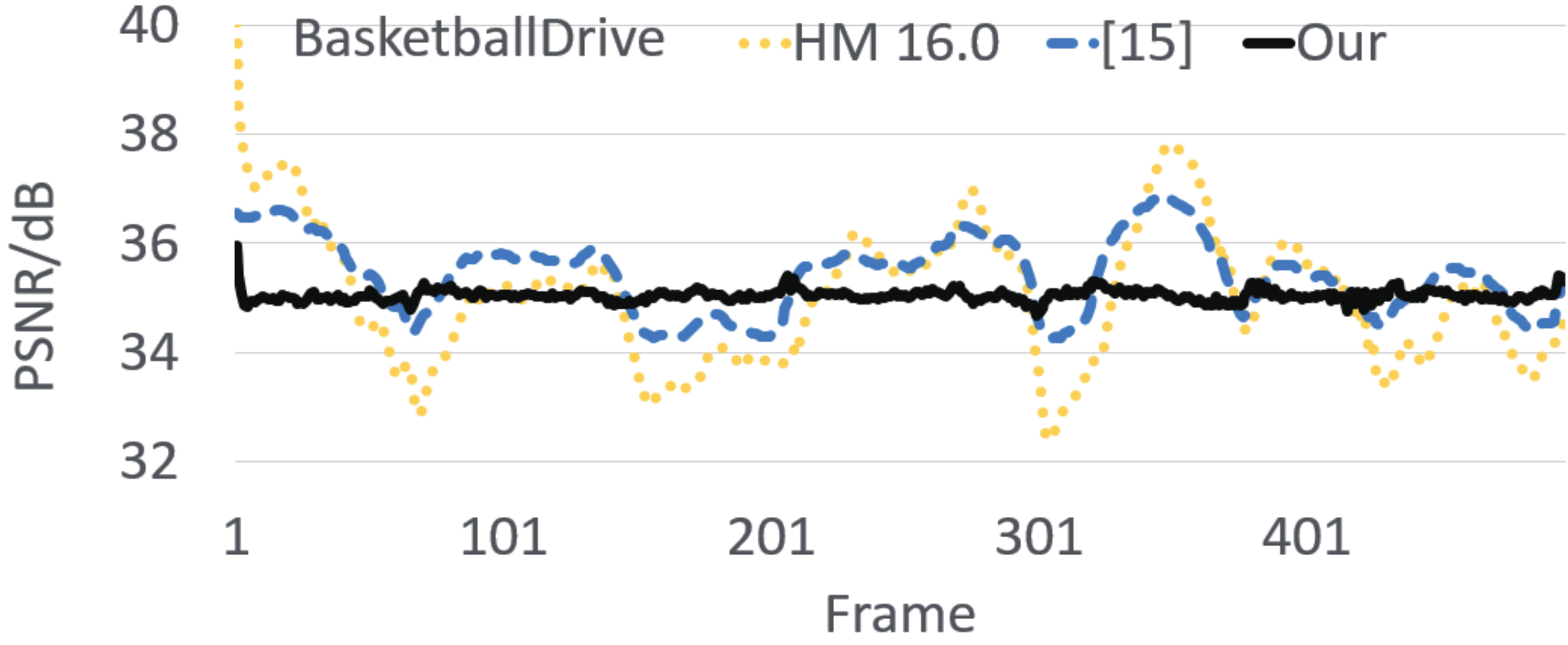}} &
\subfigure{\includegraphics[width=2.8in]{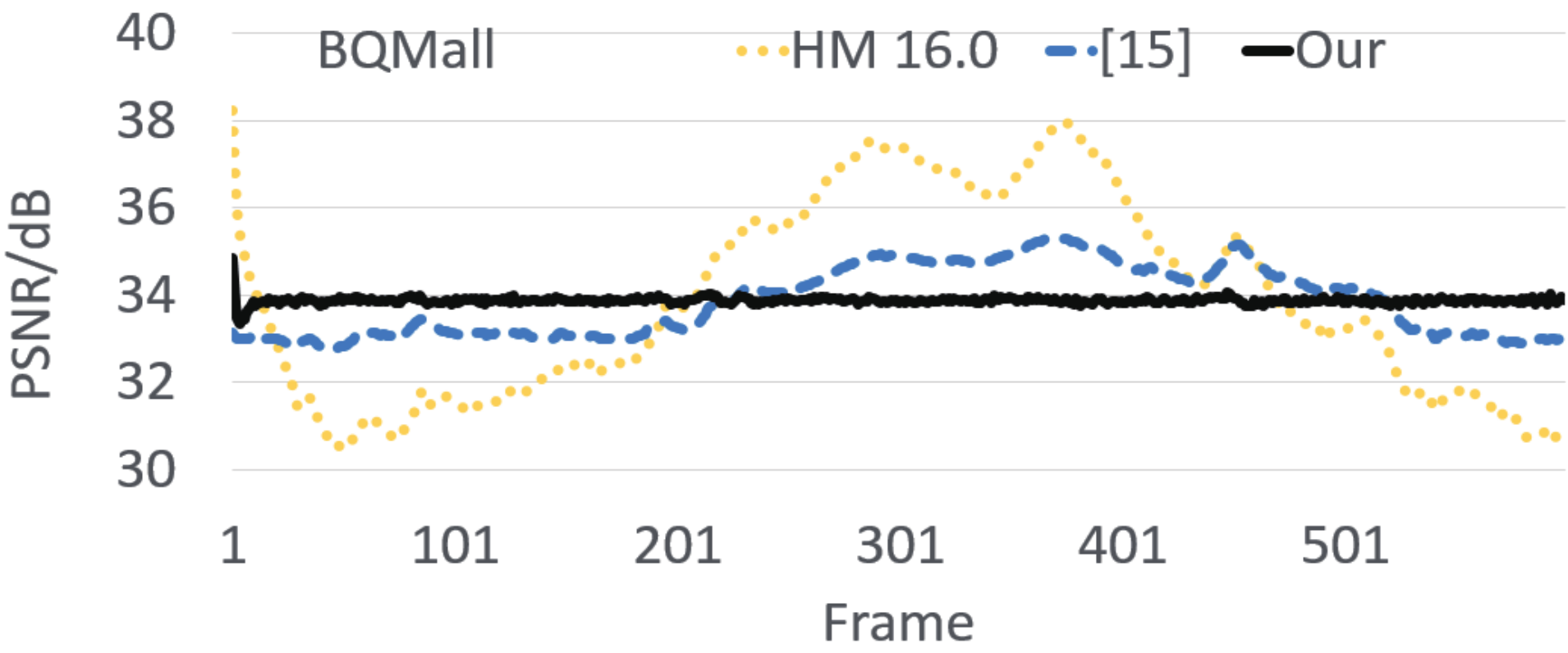}} \\
\end{tabular}
\end{center}
\caption{\label{ana}%
\footnotesize{Quality fluctuation of our PQC, conventional \cite{seo2013rate} and the default HM 16.0 methods. The quality is measured by Y-PSNR. Similar fluctuation results can be found for other encoded video sequences.}}
\end{figure}

Figure \ref{ana} illustrates the quality fluctuation of our PQC, \cite{seo2013rate} and default HM 16.0 methods, with the curves of Y-PSNR alongside frames. As can been seen from this figure, our PQC method produces the most smooth Y-PSNR curve. Similar results of quality fluctuation can be also found for other sequences, which are quantified in the 6th and 12th columns in Table \ref{result}. In these columns, the quality fluctuation is measured by the standard deviation of Y-PSNR across all frames for each video sequence. In average, our PQC method reduces the quality fluctuation for around 75\% compared with \cite{seo2013rate}. Therefore£¬ our PQC method achieves the best performance in both control error and quality fluctuation.

\begin{figure}[!h]
\begin{center}
\begin{tabular}{ccc}
\multicolumn{2}{c}{} \\[1em]
\subfigure{\includegraphics[width=2.8in]{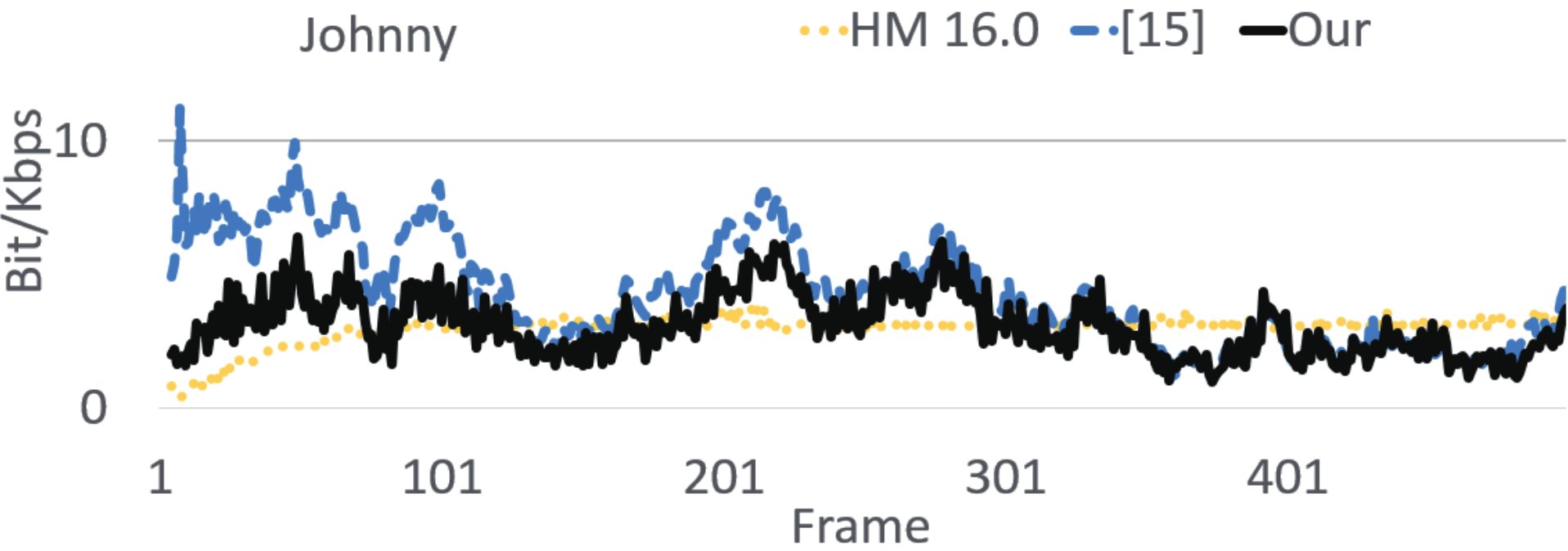}} &
\subfigure{\includegraphics[width=2.8in]{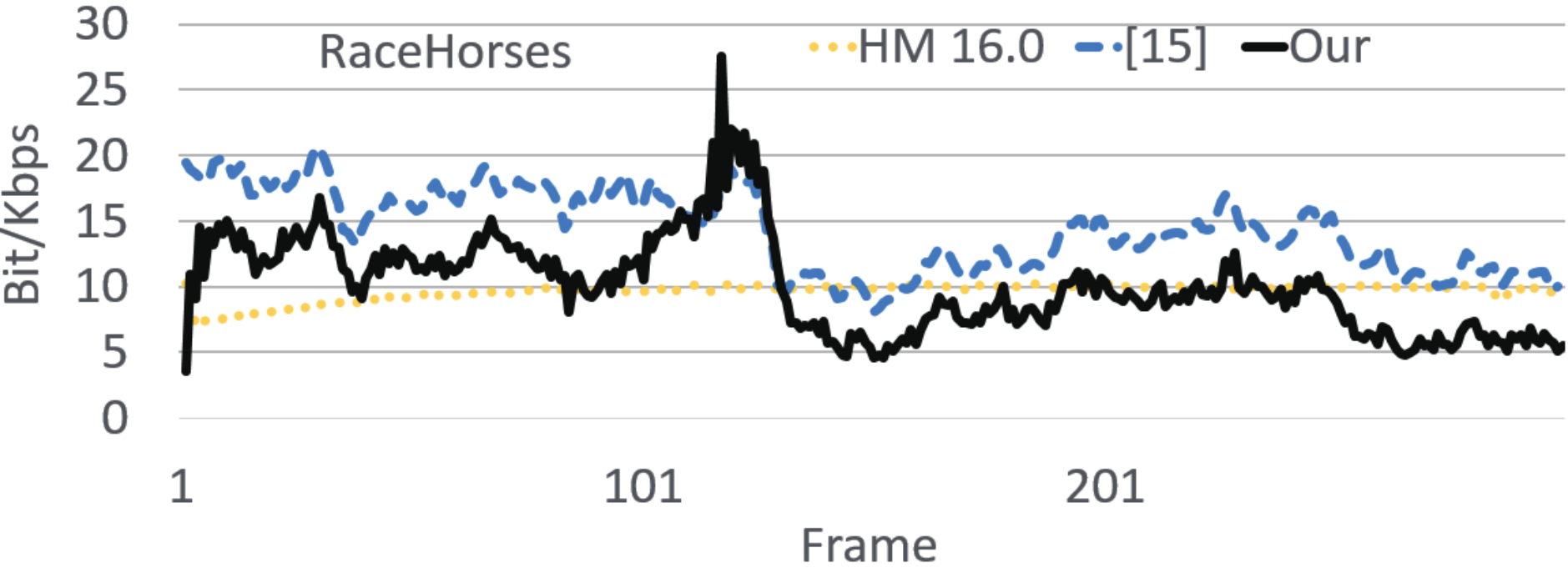}} \\
\subfigure{\includegraphics[width=2.8in]{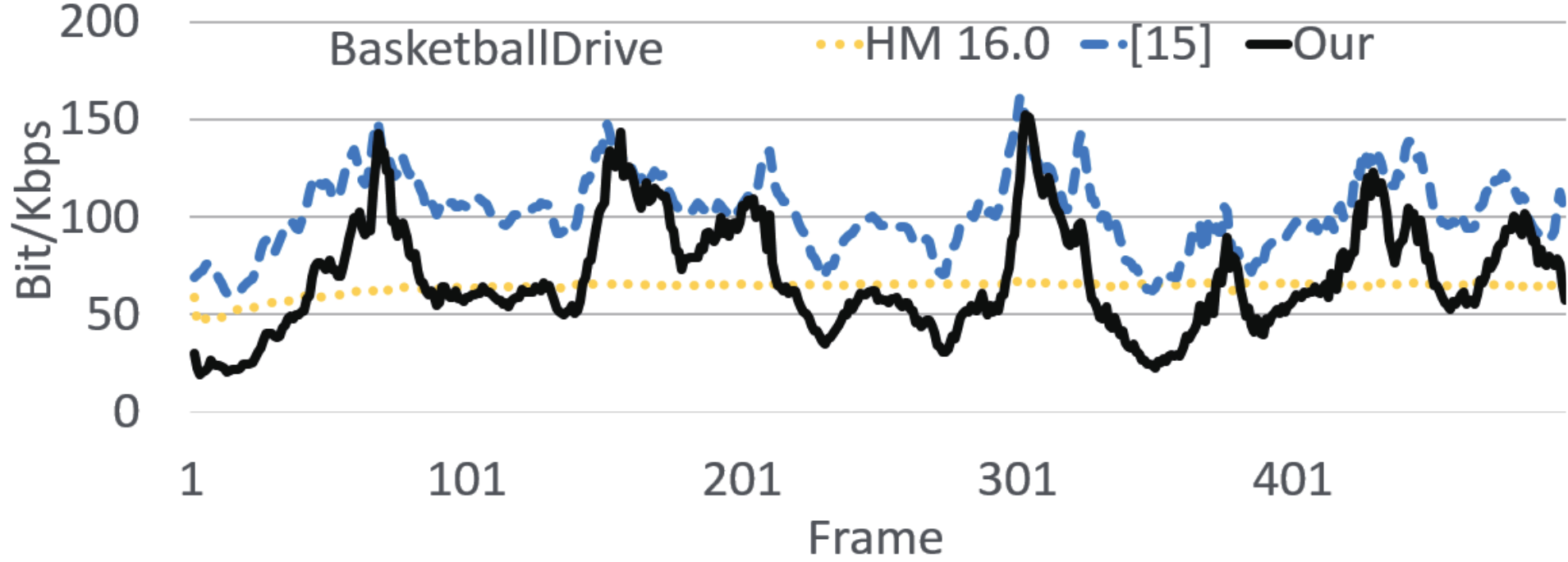}} &
\subfigure{\includegraphics[width=2.8in]{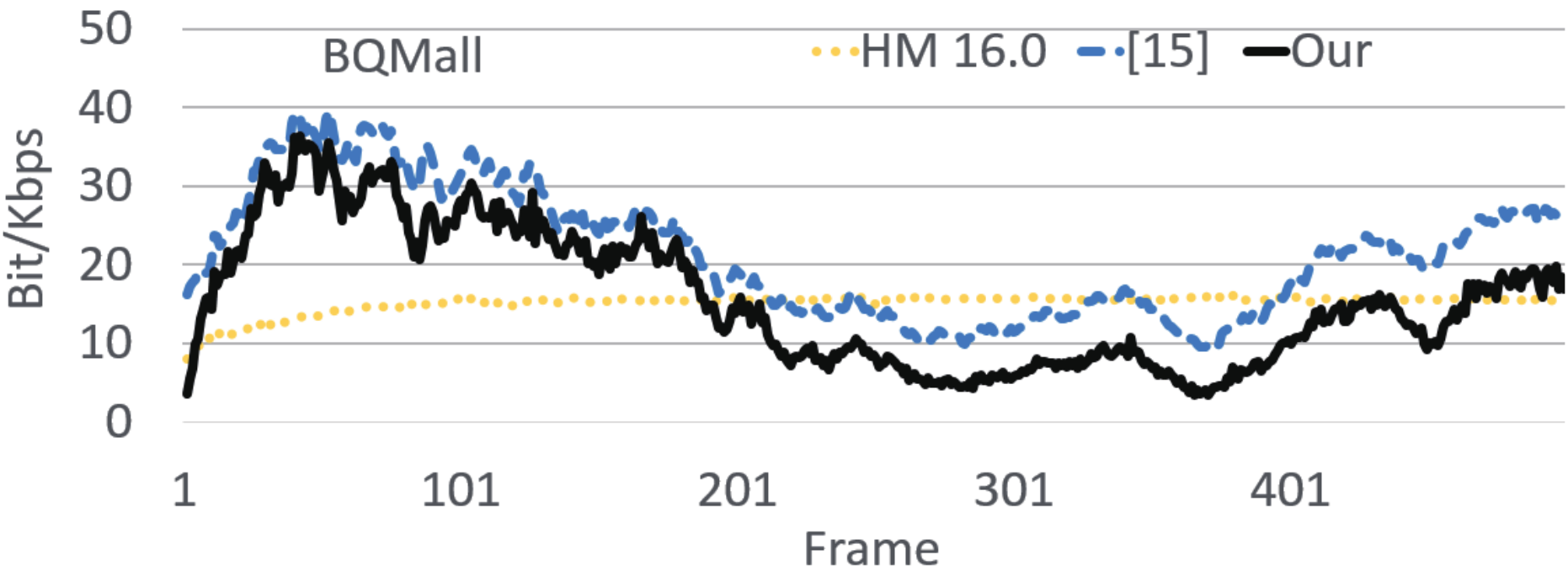}} \\
\end{tabular}
\end{center}
\caption{\label{anabit}%
\footnotesize{Bit fluctuation of our PQC, conventional \cite{seo2013rate} and the default HM 16.0 methods. Similar fluctuation results can be found for other encoded video sequences.}}
\end{figure}

\textbf{Rate-Distortion~Performance.} Here, we compare the performance of rate-distortion (R-D) performance. As reported in the 7th and 13th columns of Table \ref{result}, at similar quality, our method consumes slightly more bit-rates (around 5\% bit-rate increase) than the default HM 16.0 for most sequences. This can be seen as the cost of smoothing on quality fluctuation. However, the fluctuation is significantly reduced in our method. By contrast, the bit-rates of \cite{seo2013rate} are much higher than those of our PQC method. The reason is that \cite{seo2013rate} generally yields better quality than the target quality with more consumed bit-rates. Moreover, as shown in Figure \ref{anabit}, in order to keep a constant quality, bit-rates allocated to each frame in our method vary according to video content, thusing causing more bit-rate fluctuation when compared with HM 16.0. We also quantify the bit-rate fluctuation in Table \ref{anabit} by standard deviation of allocated bit-rates in each frame. From Table \ref{anabit}, we can observe that the bit-rate fluctuation in our method is lower than that in \cite{seo2013rate} in average. In summary, compared with \cite{seo2013rate}, our PQC method not only has a better control accuracy, but also consumes a far less bit-rate with more stable bit-rate fluctuation.

\begin{table}[tp]
\begin{center}
\caption{\label{result}%
\footnotesize{Comparison of quality control performance for HM 16.0, \cite{seo2013rate} and our method.}}
{
\renewcommand{\baselinestretch}{1}\footnotesize
\tiny{
\begin{tabular}{|c|c|c|c|c|c|c|c|c|c|c|c|c|c|}

\hline

\multicolumn{1}{|c|}{~} &
\multicolumn{1}{c|}{~} &
\multicolumn{6}{c|}{$\mathrm{QP}=32$} &
\multicolumn{6}{c|}{$\mathrm{QP}=37$} \\

\cline{3-14}

\multicolumn{1}{|c|}{Sequence} &
\multicolumn{1}{c|}{Method} &
\multicolumn{1}{c|}{Avg.} &
\multicolumn{1}{c|}{Control} &
\multicolumn{1}{c|}{Control} &
\multicolumn{1}{c|}{Quality} &
\multicolumn{1}{c|}{Bit} &
\multicolumn{1}{c|}{Bit} &
\multicolumn{1}{c|}{Avg.} &
\multicolumn{1}{c|}{Control} &
\multicolumn{1}{c|}{Control} &
\multicolumn{1}{c|}{Quality} &
\multicolumn{1}{c|}{Bit} &
\multicolumn{1}{c|}{Bit} \\

\multicolumn{1}{|c|}{~} &
\multicolumn{1}{c|}{~} &
\multicolumn{1}{c|}{PSNR} &
\multicolumn{1}{c|}{Error} &
\multicolumn{1}{c|}{Error} &
\multicolumn{1}{c|}{Fluc.} &
\multicolumn{1}{c|}{Rate} &
\multicolumn{1}{c|}{Fluc.} &
\multicolumn{1}{c|}{PSNR} &
\multicolumn{1}{c|}{Error} &
\multicolumn{1}{c|}{Error} &
\multicolumn{1}{c|}{Fluc.} &
\multicolumn{1}{c|}{Rate} &
\multicolumn{1}{c|}{Fluc.} \\

\multicolumn{1}{|c|}{~} &
\multicolumn{1}{c|}{~} &
\multicolumn{1}{c|}{(dB))} &
\multicolumn{1}{c|}{(dB)} &
\multicolumn{1}{c|}{(\%)} &
\multicolumn{1}{c|}{(dB)} &
\multicolumn{1}{c|}{(Mbps)} &
\multicolumn{1}{c|}{(Mbps)} &
\multicolumn{1}{c|}{(dB)} &
\multicolumn{1}{c|}{(dB)} &
\multicolumn{1}{c|}{(\%)} &
\multicolumn{1}{c|}{(dB)} &
\multicolumn{1}{c|}{(Mbps)} &
\multicolumn{1}{c|}{(Kbps)} \\
\hline

\multirow{3}*{BasketballDrill}
&HM &33.93&--&--&0.73&8.5&$\mathbf{1.3}$&31.59&--&--&0.72&4.4&$\mathbf{0.8}$ \\
&\cite{seo2013rate} &34.77&0.8424&2.48&0.20&11.9&4.0&32.68&1.0923&3.46&0.35&6.9&2.7 \\
&Our &33.92&$\mathbf{0.0046}$&$\mathbf{0.01}$&$\mathbf{0.13}$&9.3&3.8&31.60&$\mathbf{0.0030}$&$\mathbf{0.01}$&$\mathbf{0.11}$&4.8&2.1 \\
\hline
\multirow{3}*{BasketballDrive}
&HM &35.05&--&--&1.25&32.7&$\mathbf{3.7}$&32.73&--&--&1.31&16.6&$\mathbf{2.5}$ \\
&\cite{seo2013rate} &36.43&1.3818&3.94&0.71&51.3&25.4&34.27&1.5373&4.70&0.83&25.5&13.2 \\
&Our &35.05&$\mathbf{0.0054}$&$\mathbf{0.02}$&$\mathbf{0.18}$&34.5&27.6&32.73&$\mathbf{0.0053}$&$\mathbf{0.02}$&$\mathbf{0.20}$&17.3&13.4 \\
\hline
\multirow{3}*{BasketballPass}
&HM &33.57&--&--&2.29&4.2&$\mathbf{0.4}$&30.54&--&--&2.18&2.1&$\mathbf{0.3}$ \\
&\cite{seo2013rate} &35.94&2.3761&7.08&0.93&7.7&5.0&32.11&1.5757&5.16&1.05&3.4&2.5 \\
&Our &33.56&$\mathbf{0.0052}$&$\mathbf{0.02}$&$\mathbf{0.18}$&4.7&3.8&30.54&$\mathbf{0.0084}$&$\mathbf{0.03}$&$\mathbf{0.17}$&2.4&2.0 \\
\hline
\multirow{3}*{BlowingBubbles}
&HM &30.12&--&--&1.36&3.6&$\mathbf{0.5}$&27.27&--&--&1.31&1.5&$\mathbf{0.4}$ \\
&\cite{seo2013rate} &31.91&1.7880&5.94&0.45&5.9&5.3&28.44&1.1680&4.28&0.60&2.4&2.5 \\
&Our &30.11&$\mathbf{0.0038}$&$\mathbf{0.01}$&$\mathbf{0.16}$&3.9&3.4&27.27&$\mathbf{0.0024}$&$\mathbf{0.01}$&$\mathbf{0.17}$&1.7&1.9 \\
\hline
\multirow{3}*{BQMall}
&HM &33.87&--&--&2.26&9.4&$\mathbf{1.2}$&31.02&--&--&2.40&4.7&$\mathbf{0.7}$ \\
&\cite{seo2013rate} &35.35&1.4711&4.34&0.82&14.0&9.8&32.34&1.3220&4.26&0.93&6.8&4.7 \\
&Our &33.87&$\mathbf{0.0033}$&$\mathbf{0.01}$&$\mathbf{0.07}$&10.6&9.1&31.02&$\mathbf{0.0007}$&$\mathbf{0.00}$&$\mathbf{0.06}$&5.4&4.8 \\
\hline
\multirow{3}*{BQSquare}
&HM &30.02&--&--&0.86&3.4&$\mathbf{0.5}$&27.15&--&--&1.09&1.3&$\mathbf{0.3}$ \\
&\cite{seo2013rate} &32.36&2.3419&7.80&0.27&6.9&3.7&28.87&1.7275&6.36&0.34&2.3&1.4 \\
&Our &30.01&$\mathbf{0.0037}$&$\mathbf{0.01}$&$\mathbf{0.16}$&3.4&1.9&27.15&$\mathbf{0.0041}$&$\mathbf{0.02}$&$\mathbf{0.13}$&1.4&1.0 \\
\hline
\multirow{3}*{BQTerrace}
&HM &32.92&--&--&0.94&23.6&$\mathbf{4.2}$&30.65&--&--&1.16&8.4&$\mathbf{2.6}$ \\
&\cite{seo2013rate} &34.01&1.0927&3.32&0.66&44.6&27.9&31.93&1.2877&4.20&0.80&15.3&9.2 \\
&Our &32.92&$\mathbf{0.0017}$&$\mathbf{0.01}$&$\mathbf{0.06}$&26.9&17.9&30.65&$\mathbf{0.0010}$&$\mathbf{0.00}$&$\mathbf{0.05}$&9.3&6.7 \\
\hline
\multirow{3}*{Cactus}
&HM &34.46&--&--&0.38&26.5&$\mathbf{5.1}$&32.27&--&--&0.46&13.0&$\mathbf{3.1}$ \\
&\cite{seo2013rate} &35.18&0.7174&2.08&0.18&39.9&19.0&32.73&0.4609&1.43&0.22&18.7&9.9 \\
&Our &34.46&$\mathbf{0.0032}$&$\mathbf{0.01}$&$\mathbf{0.06}$&30.0&8.2&32.27&$\mathbf{0.0025}$&$\mathbf{0.01}$&$\mathbf{0.06}$&12.8&4.9 \\
\hline
\multirow{3}*{FourPeople}
&HM &37.12&--&--&0.71&4.4&$\mathbf{1.0}$&34.53&--&--&0.78&2.3&$\mathbf{0.7}$ \\
&\cite{seo2013rate} &38.36&1.2390&3.34&0.15&6.9&4.2&34.46&0.0695&0.20&0.14&2.7&1.7 \\
&Our &37.12&$\mathbf{0.0021}$&$\mathbf{0.01}$&$\mathbf{0.07}$&4.6&2.5&34.53&$\mathbf{0.0009}$&$\mathbf{0.00}$&$\mathbf{0.04}$&2.8&1.5 \\
\hline
\multirow{3}*{Johnny}
&HM &38.34&--&--&0.82&1.9&$\mathbf{0.6}$&35.94&--&--&1.35&1.0&$\mathbf{0.4}$ \\
&\cite{seo2013rate} &39.11&0.7768&2.03&0.32&2.6&2.2&36.62&0.6898&1.92&0.19&1.3&1.0 \\
&Our &38.34&$\mathbf{0.0022}$&$\mathbf{0.01}$&$\mathbf{0.08}$&1.9&1.1&35.94&$\mathbf{0.0016}$&$\mathbf{0.00}$&$\mathbf{0.08}$&1.0&0.6 \\
\hline
\multirow{3}*{Kimono1}
&HM &36.58&--&--&1.20&12.0&$\mathbf{6.4}$&33.95&--&--&1.60&5.9&$\mathbf{3.6}$ \\
&\cite{seo2013rate} &39.07&2.4849&6.79&0.22&12.3&15.5&36.68&2.7344&8.05&0.26&6.2&9.4 \\
&Our &36.61&$\mathbf{0.0303}$&$\mathbf{0.08}$&$\mathbf{0.12}$&12.2&8.6&34.03&$\mathbf{0.0797}$&$\mathbf{0.23}$&$\mathbf{0.14}$&6.0&3.8 \\
\hline
\multirow{3}*{KristenAndSara}
&HM &38.42&--&--&0.62&3.3&$\mathbf{0.8}$&35.86&--&--&1.01&1.7&$\mathbf{0.5}$ \\
&\cite{seo2013rate} &39.25&0.8277&2.15&0.22&5.1&3.3&36.13&0.2638&0.74&0.16&2.2&1.7 \\
&Our &38.43&$\mathbf{0.0019}$&$\mathbf{0.00}$&$\mathbf{0.06}$&3.6&1.9&35.86&$\mathbf{0.0013}$&$\mathbf{0.00}$&$\mathbf{0.05}$&2.0&1.2 \\
\hline
\multirow{3}*{ParkScene}
&HM &33.72&--&--&0.60&13.6&$\mathbf{6.3}$&31.19&--&--&0.65&5.9&$\mathbf{3.5}$ \\
&\cite{seo2013rate} &35.01&1.2838&3.81&0.18&20.7&29.6&32.23&1.0413&3.34&0.17&8.5&13.1 \\
&Our &33.72&$\mathbf{0.0086}$&$\mathbf{0.03}$&$\mathbf{0.16}$&13.7&13.4&31.19&$\mathbf{0.0058}$&$\mathbf{0.02}$&$\mathbf{0.15}$&6.0&6.4 \\
\hline
\multirow{3}*{PartyScene}
&HM &30.06&--&--&1.44&15.6&$\mathbf{6.4}$&27.18&--&--&1.38&6.6&$\mathbf{2.2}$ \\
&\cite{seo2013rate} &30.71&0.6561&2.18&0.37&18.9&14.7&27.29&0.1116&0.41&0.37&7.7&6.6 \\
&Our &30.06&$\mathbf{0.0002}$&$\mathbf{0.00}$&$\mathbf{0.23}$&16.7&12.6&27.18&$\mathbf{0.0042}$&$\mathbf{0.02}$&$\mathbf{0.10}$&7.7&6.9 \\
\hline
\multirow{3}*{RaceHorses}
&HM &32.03&--&--&1.37&3.0&$\mathbf{0.6}$&29.16&--&--&1.23&1.4&$\mathbf{0.5}$ \\
&\cite{seo2013rate} &33.66&1.6308&5.09&0.73&4.4&3.7&29.85&0.6888&2.36&0.61&1.8&2.0 \\
&Our &32.02&$\mathbf{0.0074}$&$\mathbf{0.02}$&$\mathbf{0.23}$&3.1&3.6&29.17&$\mathbf{0.0093}$&$\mathbf{0.03}$&$\mathbf{0.11}$&1.5&1.7 \\
\hline
\multirow{3}*{RaceHorses}
&HM &32.92&--&--&2.62&10.2&$\mathbf{2.1}$&29.89&--&--&2.33&4.6&$\mathbf{1.6}$ \\
&\cite{seo2013rate} &34.14&1.2177&3.70&1.13&14.9&24.8&30.40&0.5093&1.70&0.94&5.8&10.5 \\
&Our &32.92&$\mathbf{0.0085}$&$\mathbf{0.03}$&$\mathbf{0.73}$&13.4&57.5&29.90&$\mathbf{0.0133}$&$\mathbf{0.04}$&$\mathbf{0.16}$&5.8&17.5 \\
\hline
\multirow{3}*{Average}
&HM &33.94&--&--&1.22&11.0&$\mathbf{2.6}$&31.31&--&--&1.31&5.1&$\mathbf{1.5}$ \\
&\cite{seo2013rate} &35.33&1.3830&4.13&0.47&16.7&12.4&32.31&1.0175&3.29&0.50&7.3&5.8 \\
&Our &33.94&$\mathbf{0.0057}$&$\mathbf{0.02}$&$\mathbf{0.18}$&11.7&10.8&31.32&$\mathbf{0.0090}$&$\mathbf{0.03}$&$\mathbf{0.13}$&5.2&4.8 \\
\hline

\end{tabular}}}
\end{center}
\end{table}

\textbf{Computational Time.} In addition to above effectiveness evaluation, we further validate the time efficiency of our method. In our experiments,  both our method and \cite{seo2013rate} were implemented by C++ on the same computer with an Intel i7 4790k CPU and 16 GB DDR4 memory. Then, the computational time of both our method and \cite{seo2013rate} was record for each video sequence. We found that our method averagely increases $2.3 \mu s$ per frame, whereas \cite{seo2013rate} takes up $31.4 \mu s$ per frame. Thus, we can conclude that our method is much faster than \cite{seo2013rate}, and it hardly increases the computational time of video coding. The fast speed of our method is mainly due to the fact that the quality control of our method, achieved by \eqref{final-policy-function-inter} and \eqref{final-policy-function-intra}, only requires 10 times of floating-point addition operations and 10 times of floating-point multiplication operations.

\vspace{-1.0em}
\section{Conclusion}
\vspace{-1.0em}
In this paper, we have proposed a new method, called PQC method, to optimally control the quality of video coding. First, we established a formulation for optimal quality control, which considers both error and fluctuation of quality control in video coding. For the established formulation, the relationship between $\mathrm{QP}$ and control error was modelled in our PQC method. Then, a PID-based solution was developed to solve our quality control formulation, based on the modelled relationship between $\mathrm{QP}$ and control error alongside encoded frames. Since our PQC method only alternates $\mathrm{QP}$ values of encoded frames for achieving desirable control error and quality fluctuation, it can be seen as an encoder-free method. We further implemented our PQC method in the latest HEVC encoder, i.e., HM 16.0. The experimental results show that the control error and quality fluctuation of our PQC method are much better than those of the conventional method for HEVC.

\Section{References}

\end{document}